\documentclass[a4paper,10pt]{article}
\usepackage{graphicx}
\usepackage{latexsym,amsmath,verbatim}

\newcommand{\bea}{\begin{eqnarray}}
\newcommand{\eea}{\end{eqnarray}}

\title{Mechanisms of  Self-Organization and Finite Size Effects in a Minimal Agent Based Model}

\author{V. Alfi$^{1,2}$, M. Cristelli$^1$, L.Pietronero$^{1,3}$, A. Zaccaria$^1$\\\\
$^1$ \small Universit\`a ``La Sapienza'', P.le A. Moro 2, 00185, Roma, Italy\\
$^2$ \small Centro ``E. Fermi'', Compendio Viminale, 00184, Roma, Italy\\
$^3$ \small  ISC-CNR, Via dei Taurini 19, 00185, Roma, Italy
}

\begin{document}

\maketitle

\begin{abstract}
We present a detailed analysis of the self-organization phenomenon in which
the stylized facts originate from finite size effects with respect to the number 
of agents considered and  disappear in the limit of an infinite population.
By introducing the possibility that agents can enter or leave the market
depending on the behavior of the price,
it is possible to show that the system self-organizes in a regime with a finite number of agents which
corresponds to the stylized facts.
The mechanism to enter or leave the market is based on the idea that a too stable market is unappealing for traders while the presence of price movements attracts agents to enter and speculate on the market.
We show that this mechanism is also compatible with the idea that agents are scared by a noisy and risky market at shorter time scales.
We also show that the mechanism  for  self-organization is robust with respect to 
variations of the exit/entry rules and that the attempt to trigger the system to self-organize in a region without stylized facts leads to an unrealistic dynamics.
We study the self-organization in a specific agent based model but we believe that the basic ideas should be of general 
validity.
\end{abstract}

\section{Introduction}

In this paper we discuss in detail the  self-organization phenomenon which 
emerges in the minimal Agent Based Model (ABM) we have introduced in~\cite{paperNP,paperoI,paperoII}.
This model has the great advantage to be
simple, mathematically well posed and able to reproduce the stylized facts (SF), i.e. the
empirical evidences of real markets~\cite{mantegnabook,bouchaud,cont}.
In this respect it can be considered 
a ``workable model''  for which analytical approaches are possible in some cases
 and it can be easily modified to introduce variants and elements of realism.
\\
We consider a population of heterogeneous interacting agents whose 
strategies are divided in two main classes: fundamentalist and chartist.
Fundamentalist agents believe that the market is stable and fluctuates around a fair value, the fundamental price $p_f$, that they 
estimate by standard economic arguments. The fundamentalist strategy acts in a way to drive 
the price towards the fundamental value.
Chartist agents instead try to detect local trends in the market by evaluating the price history.
They try to speculate betting on these trends and so they contribute to the formation of bubbles and crashes.
Agents can change their strategy during the dynamics by following some personal considerations or by imitating other agents behavior (herding).
It is possible to see that, by considering a given finite number of agents, the dynamics shows
an intermittent behavior. This means that  
the market is assumed to be dominated by fundamentalists at large times
but bursts of chartists can appear spontaneously leading to high volatility.
In principle the basic assumption of large time stability can be removed 
if one would like to consider particularly turbulent situations but in the present study we will
only refer to the simple case mentioned before.
\\
The intermittent behavior is present only for a particular value of the number of agents
and it disappears in the limit of infinite agents.
This phenomenon has been already observed in other similar models
but it has been interpreted as a negative element.
Here we present a completely different perspective by showing that 
the finite number of agents necessary to produce the SF is not an artificial feature of the model
to be eliminated.
On the contrary this finite size effect results to be the natural outcome of a process of
self-organization.
\\
The basic concept is that the self-organization can be triggered by leaving the agents
the possibility to enter or exit the market following a mechanism based on a feedback on the
price behavior.
This mechanism encourages agents to enter the market if they perceive an interesting movement
in the price.
On the other hand a very stable market where nothing happens is not appealing for speculators
who are likely to abandon such a market. 
From an economic point of view a very stable market can be attractive for some
particular agents who only look for the conservation of their wealth,
but these will not contribute to the SF and are irrelevant with respect to our
discussion.
\\
This dynamics for the agents is implemented by introducing two suitable thresholds which agents consider to decide to enter or leave the market by comparing them with the price movement.
By considering 
various initial situations with different starting number of agents
we can observe that the resulting dynamics stabilizes around a finite number of agents
which is the one corresponding to  the SF.
This phenomenon corresponds to the self-organization of the system in a state dominated by an intermittency due to finite size effects.
In this respect it is not a case of self-organized criticality~\cite{perbak,jensen} but rather of self-organized intermittency.
\\
The fact that real agents can be scared by a market that is too volatile may appear
at first sight problematic with respect to our criteria to enter or exit the market.
This point requires a clarification of the time scales involved.
For a market to be attractive, there has to be some price movement at a relatively 
long time-scale corresponding to the operation performed.
On the other hand volatility at shorter time-scales
may indeed appear as a disturbance for an agent.
We have checked this point by analyzing the volatility also at short time-scale
and considering this as a discouraging element for the agents.
The general result is that the introduction of this additional and realistic element
does not modify appreciably the phenomenon of self-organization discussed before.
We have also checked the robustness of the self-organization mechanism along various directions.
For example it is possible to see that 
if one tries to force the system to self-organize in a state without the SF
one meets unrealistic scenarios.
\\
 In summary we propose a simple mechanism  for the agents' dynamics which is able to explain why real
 markets self-organize in a state corresponding to the SF. 
 This mechanism is based on the idea that speculative agents dislike a too stable market and prefer
 to bet on price movements which they interpret as opportunities to exploit 
 following their strategies.
This mechanism is stable and does not contradict the natural fear of real traders to enter
in risky, highly fluctuating markets with respect to shorter time-scales.
\\
The paper is organized in the following way.\\
In section~\ref{sec:2} we give a schematic description of the ABM introduced in \cite{paperNP,paperoI,paperoII}.
\\
In section~\ref{sec:3}  we discuss the basic mechanisms which lead to the self-organization towards
the region with intermittent dynamics and SF.
We also consider some variants and check the stability of the mechanism.
\\
In section~\ref{sec:4} we propose a more realistic generalization of the model with two temporal horizons to define the entry/exit strategies of the agents. In this way we can include the tendency of real traders to be scared by a noisy market and show that this does not interfere with the self-organization.
\\
In section~\ref{sec:5} the conclusions are drawn and we also outline some possible perspectives
of the present study.

\section{The Minimal Agent Based Model in a Nutshell}
\label{sec:2}

The mathematical framework
which we consider to study the self-organization mechanism
is the minimal  ABM we have introduced in \cite{paperNP,paperoI,paperoII}. This ABM is composed of $N$ agents that can be chartists ($N_c$) or fundamentalists ($N_f$) and clearly $N=N_c+N_f$. 
The novelty of this model, which makes it a workable tool
to consider a variety of questions, is a simplification of the elements to those which are strictly essential.
In addition the equations for the dynamical evolution are mathematically well posed and general 
without any {\itshape ad hoc} assumption ~\cite{Lux:1999,lebaron,giardina}.
\\
The chartists are recognized as destabilizing agents and an efficient way to describe them is represented by the potential method introduced in
~\cite{vale2,vale3,taka1,taka2}. 
The action of these investors can be described, in the simplest case, in terms of a repulsive force proportional to the distance between the current price $p(t)$ and a suitable moving average $p_M(t)=M^{-1}\sum_{j=0}^{M-1}p(t-j)$. 
\\
On the other hand the fundamentalists believe in a fundamental price $p_f$ and bet on a reverting trend towards this value. A simple way to mimic their action is to define an AR(1) process where $p_f$ plays the role of an attractor. 
\\
We now assume that the price formation can be described in term of a linearized Walras' mechanism (i.e. $dp/dt=ED$ where $ED=\text{excess demand}$) and the complete equation of price dynamics is consequently,
\begin{equation}
p(t+1)=p(t)+\sigma\xi(t)+\frac{1}{N}\bigg[\sum_{i=1}^{N_c}\frac{b_i}{M_i-1}\big(p(t)-p_{M,i}(t)\big)+\sum_{j=1}^{N_f}\gamma_j\big(p_f-p(t)\big)\bigg].
\end{equation}
For the sake of simplicity we set $\gamma_j=\gamma\,\,\forall j$, $b_i=b\,\,\text{and}\,\,M_i=M\,\,\forall i$ which are respectively the strength of fundamentalist action, of the chartist action and the memory of the moving average. The term $\sigma\xi$ is the white noise whose amplitude is fixed by $\sigma$.
\\
The key element of the model is the dynamics of the evolution of the strategies. Here we use the simplified version of the probability of switching a strategy that models only an asymmetric herding effect. 
This asymmetry guarantees that fundamentalists will prevail at very long times
(for a more detailed discussion see \cite{paperNP,paperoI}),
\begin{equation}
\;\;P_{cf}=B(1+\delta)(K+\frac{N_f}{N})\
\label{eq:dynamics1}
\end{equation}
\vspace*{-0.2cm}
\begin{equation}
P_{fc}=B(1-\delta)(K+\frac{N_c}{N})
\label{eq:dynamics2}
\end{equation}
where $P_{cf}$ and $P_{fc}$ are respectively the probability of becoming fundamentalist being chartist and the probability of becoming chartist being fundamentalist. The parameter $K$ prevents the dynamics to be captured indefinitely by the absorbing states $N_c=0$ and $N_f=0$. We set $K=r/N$ with $r<1$ in order to be able to vary the number of agents $N$ without quitting the region of parameters in which the probability density function of the population is bimodal~\cite{paperNP,paperoI,kirman:1993,alfalux}. \\
Further and more exhaustive discussions about the intermittent behavior of the population dynamics, the origin of the volatility clustering and in general about the statistical properties of the model can be found in~\cite{paperNP,paperoI,paperoII}.

\section{Basic mechanism for the self-organization}
\label{sec:3}

\subsection{Self-organized intermittency}\label{sec:3.1}

\begin{figure}[ht!]
\begin{center}
\includegraphics[scale=0.40]{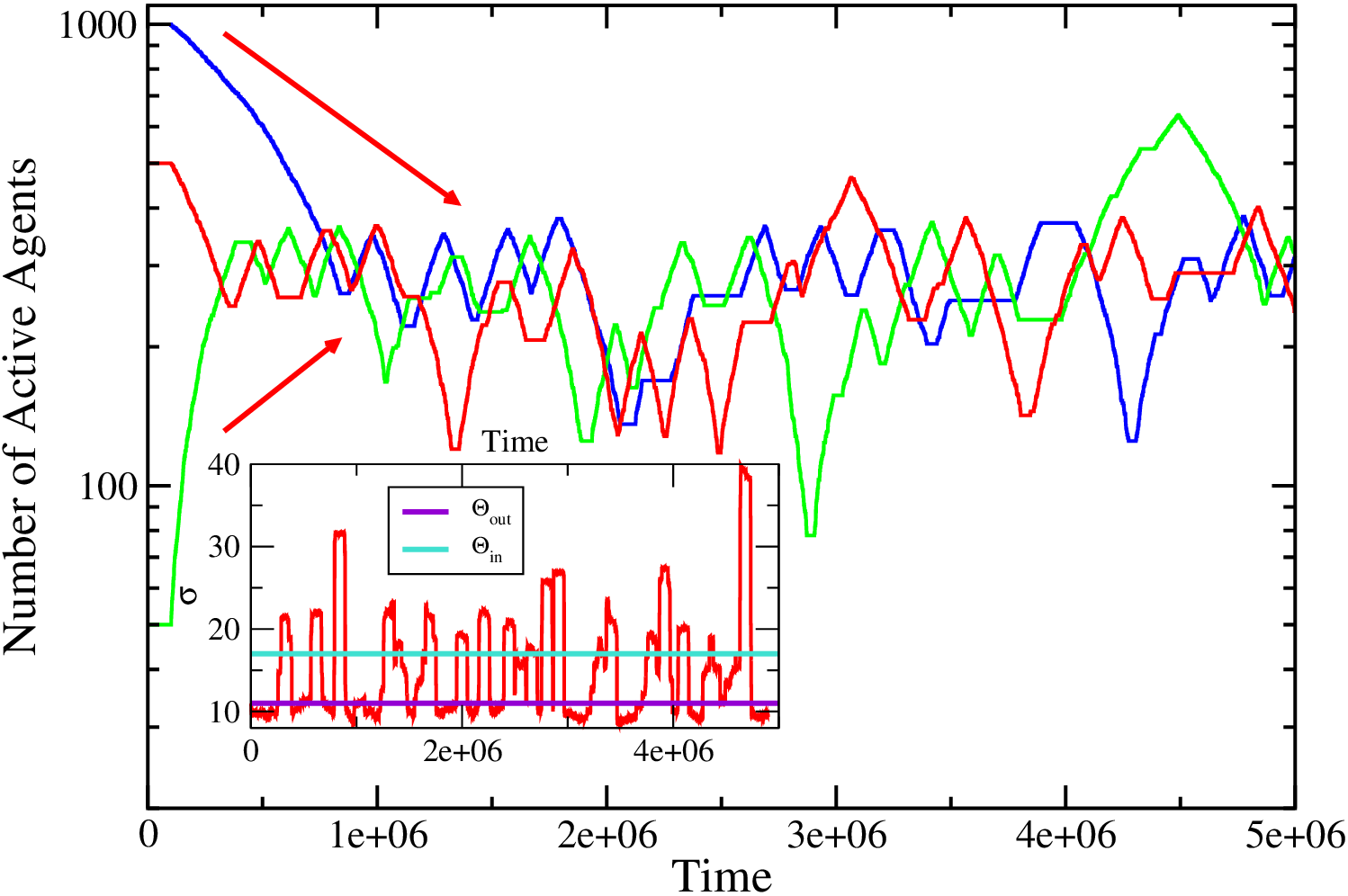}
\caption{Self-organization to the quasi-critical state. Three populations with different starting number of agents ($N=50$ green line, $N=500$ red line and $N=1000$ blue line) evolve in time, following the threshold rule, towards the quasi-critical intermittent state which corresponds to the SF. Starting from a large population ($N=1000$) agents exit the market because it is too stable. The opposite happens
when the starting number of agents is small ($N=50$). In this case agents enter the market to exploit the
profit opportunities they detect. When the populations starts from $N=500$
the number of agents fluctuates around an almost stable value which is the one of the intermittent dynamics leading to the SF. In the inset the time fluctuations of the estimator $\sigma(t,T)$ are shown together with the thresholds ($\Theta_{in}$ and $\Theta_{out}$) used in the simulations.
We can see the intermittent dynamics of the volatility $\sigma(t,T)$.}
\label{fig:soi}
\end{center}
\end{figure}

Eqs.~\ref{eq:dynamics1} and \ref{eq:dynamics2} define the dynamics of $N_c$ and $N_f=N-N_c$. If we consider
the variable $x=N_c/N$ it is possible to see that the distribution
of $x$ depends explicitly on the value of the total number of agents $N$ active in the market~\cite{paperNP,paperoI}.
In particular when $N$ is very large 
the transition probability from fundamentalist to chartist becomes asymptotically small
and essentially the system becomes frozen in the fundamentalist state.
The resulting price dynamics is then extremely stable
due to the stabilizing effect of fundamentalists.
Clearly such a state does not show the anomalous fluctuations corresponding to the SF.
This means that in the limit of an infinite size system $N$ the resulting dynamics looses the interesting properties which lead to the formation of the SF.
This is different from what happens in the majority of physical models where the interesting (critical)
phenomena appear in the thermodynamic limit.
The opposite happens in the limit of very small $N$ where the population of agents
undergoes very fast changes of strategy and the resulting dynamics is too {\itshape schizophrenic}.
Only for an intermediate number of agents (the specific value depending on the other model parameters)
one can recover the intermittent dynamics of the changes of strategy which leads to the SF.
Of course this situation leads to the basic question of understanding the driving mechanism
which makes real markets self-organize in the intermittent state with a finite number of agents.
\\
The first consideration in this respect is that the number of agents $N$
should be itself a fluctuating variable and not a fixed parameter of the model.
This implies the identification of a mechanism which rules the decision of agents to enter or leave the market
depending on the price behavior.
\\
The idea is that traders are usually attracted to enter in the market if they 
detect an interesting signal 
in the price behavior which usually is an appreciable movement of the price.
Otherwise, if the market is too stable, no gain opportunities appear and traders leave the market.
We would like to stress that both chartist and fundamentalist agents described in our model are a kind of speculative traders in the sense that they try to profit by betting in future movements of the price.
The behavior of extremely conservative investors 
who appreciate a completely stable situation
is not part of this scheme because they do not contribute to the fluctuations leading to the SF.
\\
The price movements which are attractive to agents are estimated in our model by considering a 
long-term ($T\approx1000$) estimation of the standard deviation of the price,
\begin{equation}
\sigma(t,T)=\sqrt{\frac{1}{T-1}\sum_{i=0}^{T-1}(p(t-i)-\bar{p})^2}.
\label{eq:sigma}
\end{equation}
This quantity $\sigma$ is the one agents consider to decide to enter or leave the market.
If $\sigma$ is small agents leave the too stable market where no profit opportunities appear.
Otherwise if $\sigma$ is sufficiently large then significant movements in the price behavior are expected 
and agents enter the market.
This situation can be described in terms of the two threshold values $\Theta_{in}$ and $\Theta_{out}$.
In particular agents will (in a probabilistic way) enter the market if $\sigma(t,T)>\Theta_{in}$
and leave the market if $\sigma(t,T)<\Theta_{out}$:
\begin{equation}
\left\{
\begin{array}{cc}
\mbox{enter if} & \sigma(t, T)>\Theta_{in}\\
\mbox{exit if} & \sigma(t, T)<\Theta_{out}
\label{eq:2thresh}
\end{array}
\right.
\end{equation}
In Fig.~\ref{fig:soi} we can observe the phenomenon of self-organization. 
Starting from a small value of
$N$ $(N = 50)$ the large price movements will attract more agents and $N$ increases. 
Starting instead from a large populations of agents $(N = 5000)$ the opposite happens and 
$N$ decreases. For $N=500$ we have a relatively
stable situation. The self-organization to the intermittent state which leads to the SF
corresponds therefore to the fact that this is the attractive fixed point for the dynamics of the agents related to their threshold strategies. The fact that this occurs in our model for $N=500$ depends on the specific parameters we have adopted but clearly the phenomenon of self-organization is completely general
and robust. In principle, by changing the parameters, one could have the intermittent state at different values of $N$.
In the inset of Fig.~\ref{fig:soi} we have plotted the intermittent behavior of the estimator $\sigma$ as a function of the time compared with the two threshold $\Theta_{in}$ and $\Theta_{out}$.
We can call this attractive intermittent state ``quasi-critical'' to distinguish it to the usual critical 
state of statistical physics model.
\\
Along this line of reasoning we can make two comments with respect to real market:
\\
i) different markets may correspond to very different  parameters in our model. For each set of this
parameter there would be a self-organization to  a quasi critical value N* leading to the SF.
This permits to explain that the number of agents can be different in different markets
still they all lead to SF with similar properties.
However, this intermittent properties are due to finite size effects and for this reason
a strict universality should not be expected;
\\
ii) in our model the total  number of agents $N$ has a very precise mathematical meaning.
It corresponds to the number of independent interacting variables in the model.
The interpretation of an effective $N$ in a real market is therefore a subtle problem which
requires a careful analysis.
The herding mechanism induces a tendency of agents to behave similarly
but this is not strictly compulsory. In reality if a group of agents, for whatever reason, act coherently
in the market they cannot be considered as mathematically different agents but are essentially a 
single agent. For this reason the estimation of the effective number
of independent agents in a real market 
represents a very interesting and important problem.

\subsection{Other possible rules}

In section \ref{sec:3.1} we have seen that,  by fixing 
the thresholds $\Theta_{in}$ and $\Theta_{out}$ in a region of values 
of the volatility $\sigma$ which is intermediate with respect to the two limit cases
$\sigma_{N=5000}$ and $\sigma_{N=50}$,
the system self-organizes in the intermittent state corresponding to an intermediate number of agents 
$N^*\simeq 500$.
\\
In doing this we have chosen the threshold values to correspond approximatively to the region 
of the fluctuations leading to the SF.
This may induce the idea that by choosing different values of $\Theta_{in}$ and $\Theta_{out}$
one may force the system to self-organize to any preselected state, not necessarily
the one corresponding to the SF.
We are going to see that it is not the case because by choosing unreasonable thresholds' region 
the system does not reach an interesting or unique self-organized state.
\begin{figure}[t!]
\begin{center}
\vspace{1cm}
\includegraphics[scale=0.40]{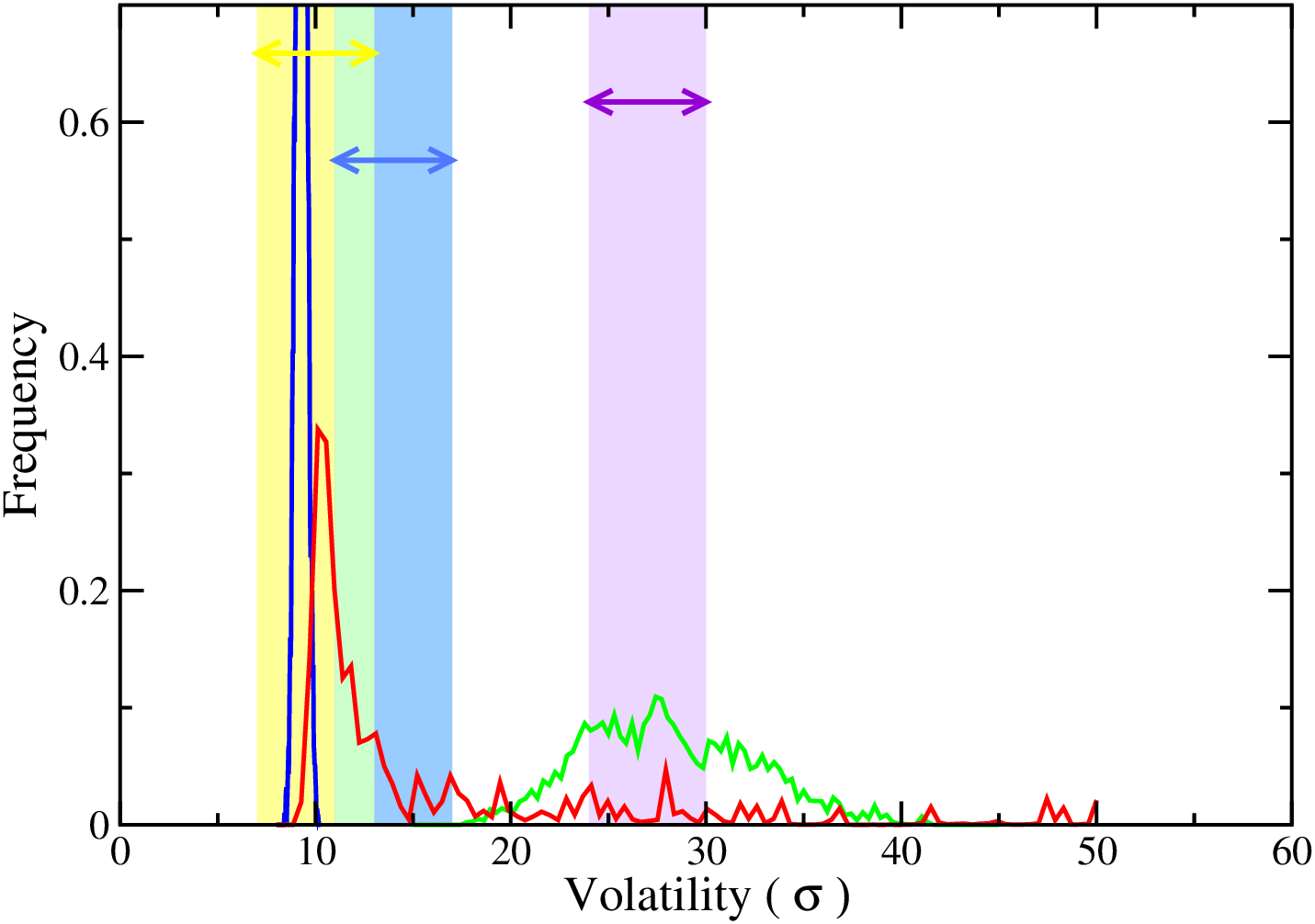}
\caption{Histograms of the volatility $\sigma$ for different populations with fixed number of agents $N$. The picture refers to three populations with $N$ equal to $50$ (green line), $500$ (red line) and $5000$ (blue line). The histograms referring to $N=50$ and $N=5000$ are almost symmetric with the difference that the first is broad and the second very sharp. The histogram for $N=500$ is asymmetric and has tails which extends for very large values of the volatility $\sigma$, even larger than the 
ones of the $N=50$ histogram.
By considering this plot, we have identified three different regions
(indicated in the picture with different colors), delimited
by different values of $\Theta_{in}$ and $\Theta_{out}$, to trigger the self-organization towards different values of $N$. 
We will see that only by choosing
a suitable region centered on the maximum of the histogram which refers to $N=500$ one can obtain a market dynamics which self-organizes towards a stable value.
}
\label{fig:3regions}
\end{center}
\end{figure}
In Fig.~\ref{fig:3regions} we have plotted the histograms of the volatility $\sigma$ corresponding
to three different populations with a fixed number of agents $N$ with values $50,500,5000$.
We can see that the volatility of the $N=5000$ population is very sharp and it is picked
around a small value of $\sigma$. In the case of $N=50$ the histogram is broader and has a 
maximum on a very high level of volatility.
The situation is different in the intermediate case of $N=500$ where
the distribution in very broad and asymmetric. It is picked on small values of volatility but the tails
reach very high values, much more than the $N=50$ population.
The reason for the high values of price fluctuations of the intermediate case ($N=500$)
with respect to the extreme case ($N=50$) is the following.
A very large price fluctuation corresponds to a situation in which 
the chartists action can develop for a certain time. In the highly fluctuating regime ($N=50$)
the life time of chartist action is too small for this to happen.
On the contrary for the intermediate case chartist fluctuations are more rare but when they 
happen they may last for a longer time.
\\
We now consider three different 
possibilities for the thresholds values
of $\Theta_{in}$ and $\Theta_{out}$. 
These regions are evidenced in Fig.~\ref{fig:3regions} with different colors.
The first region is centered on low values of volatility, the second
on high values and the third on intermediate values. By choosing this 
last region, that is the one used in section~\ref{sec:3.1}, the system self-organizes
in the intermittent state which corresponds to a fluctuating intermediate number of agents $(N\simeq500)$.
We are going to see that unrealistic anomalies occur if
one chooses the other two regions.
\begin{figure}[htbp]
\begin{center}
\vspace{1cm}
\includegraphics[scale=0.23]{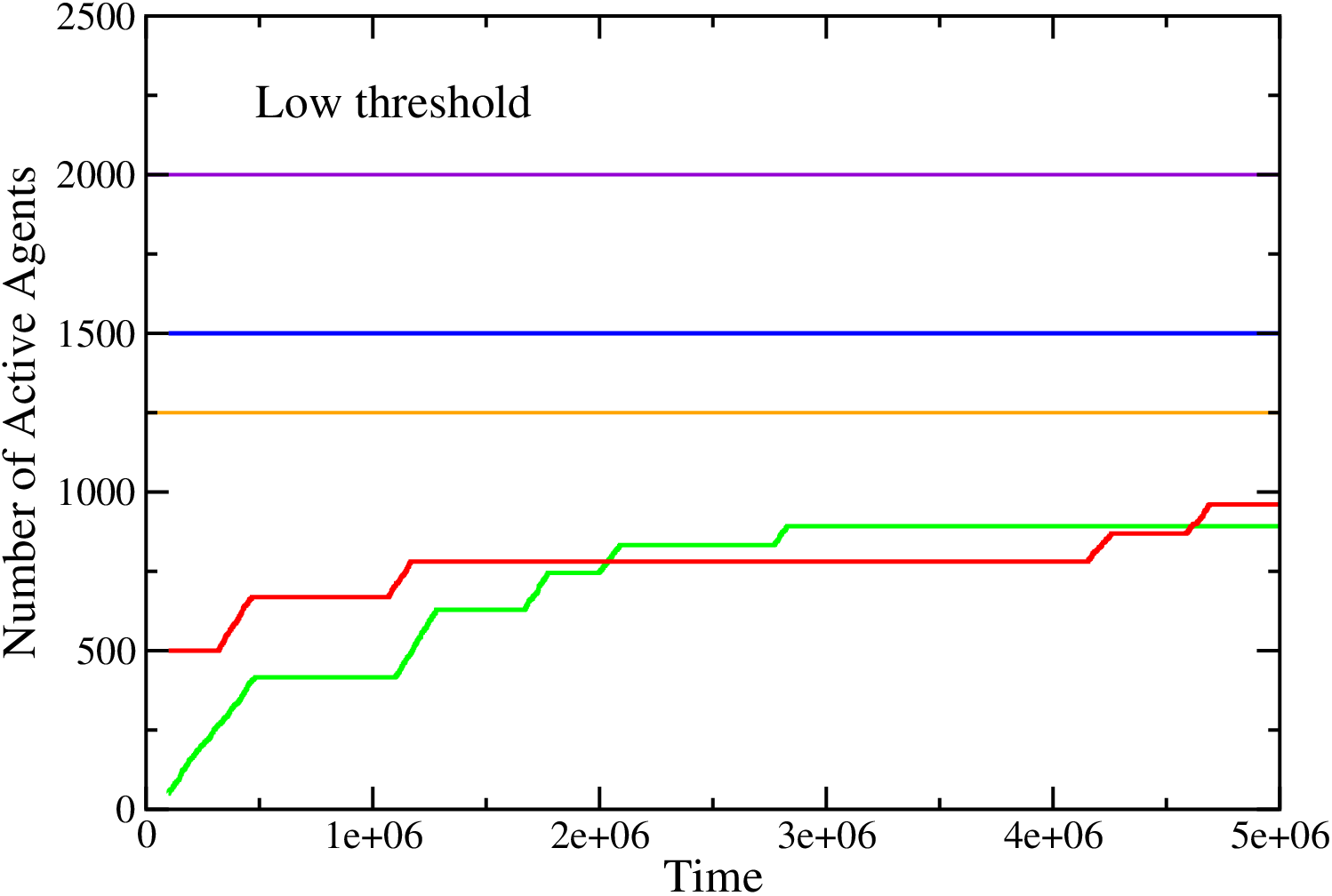}
\includegraphics[scale=0.23]{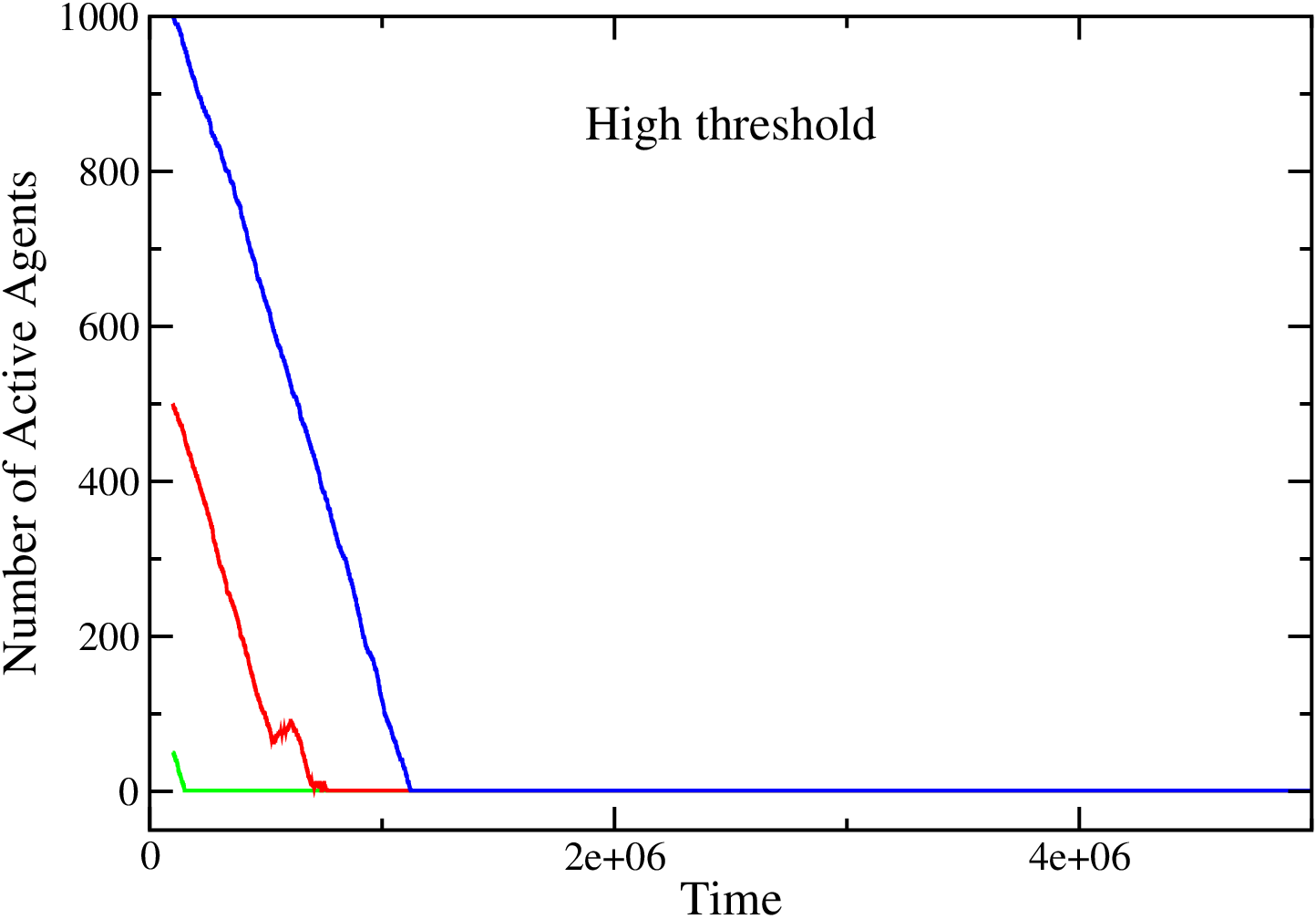}
\caption{Self-organization using other rules. In this plot we have analyzed
the self-organization phenomenon using
values of $\Theta_{in}$ and $\Theta_{out}$ which defines regions which are centered on low (left plot) and high (right plot) values of volatility. 
We can see that by choosing these regions the self-organization phenomenon is no more observed. In the left panel we can see that the number of agents $N(t)$ is 
an always increasing function which becomes a constant when $N(t)$ exceeds
a certain value.
The opposite happens in the right panel where $N(t)$ decreases step by step
going towards the unrealistic situations with $N=0$.
}
\label{fig:highandlow}
\end{center}
\end{figure}
When the region defined by $\Theta_{in}$ and $\Theta_{out}$ is centered on low volatility values, and 
the starting number of agents is small,
the system size grows until it reaches a number of agents
which leads to an average value of the volatility $\sigma$ which is inside the region considered.
Then the fluctuations from this average value are so small that the system is actually locked
in a certain (high) value of the number of agents. The dynamics corresponding to this situation
does not display the SF anymore.
Of course when the starting number of agents is very large its average volatility level 
is always inside the considered region and the system size is constant in time.
The dynamics corresponding to a low volatility centered region is shown in Fig.~\ref{fig:highandlow}(left)
and it does not lead to the phenomenon of the self-organization.
\\
On the contrary, as shown in Fig.~\ref{fig:highandlow}(right), when the region defined by the thresholds is centered on very high values
of volatility the system size rapidly drops down because the system has an average 
volatility which is always smaller than the threshold considered to enter the market.
In this way the system collapses to the unrealistic situation of zero-agent population where
the price fluctuations are only due to the random noise term.
\\
Therefore this study shows that the phenomenon of self-organization and the presence of stylized 
facts are intrinsically linked and one cannot force the system to self-organize to a reasonable dynamics which does not lead to the SF.
For example this forbids the possibility of  self-organization associated to a random walk dynamics (associate to 
the SF).

\subsection{Further simplification of the mechanism for the self-organization}

In this subsection we consider a simple modification of the mechanism described in
section~\ref{sec:3.1} to obtain a self-organized market.
The idea is that agents enter the market if the volatility  is larger
than a certain threshold $\Theta_{in}$ while they leave the market 
if this volatility is smaller than the threshold $\Theta_{out}$.
\begin{figure}[t!]
\begin{center}
\vspace{1cm}
\includegraphics[scale=0.40]{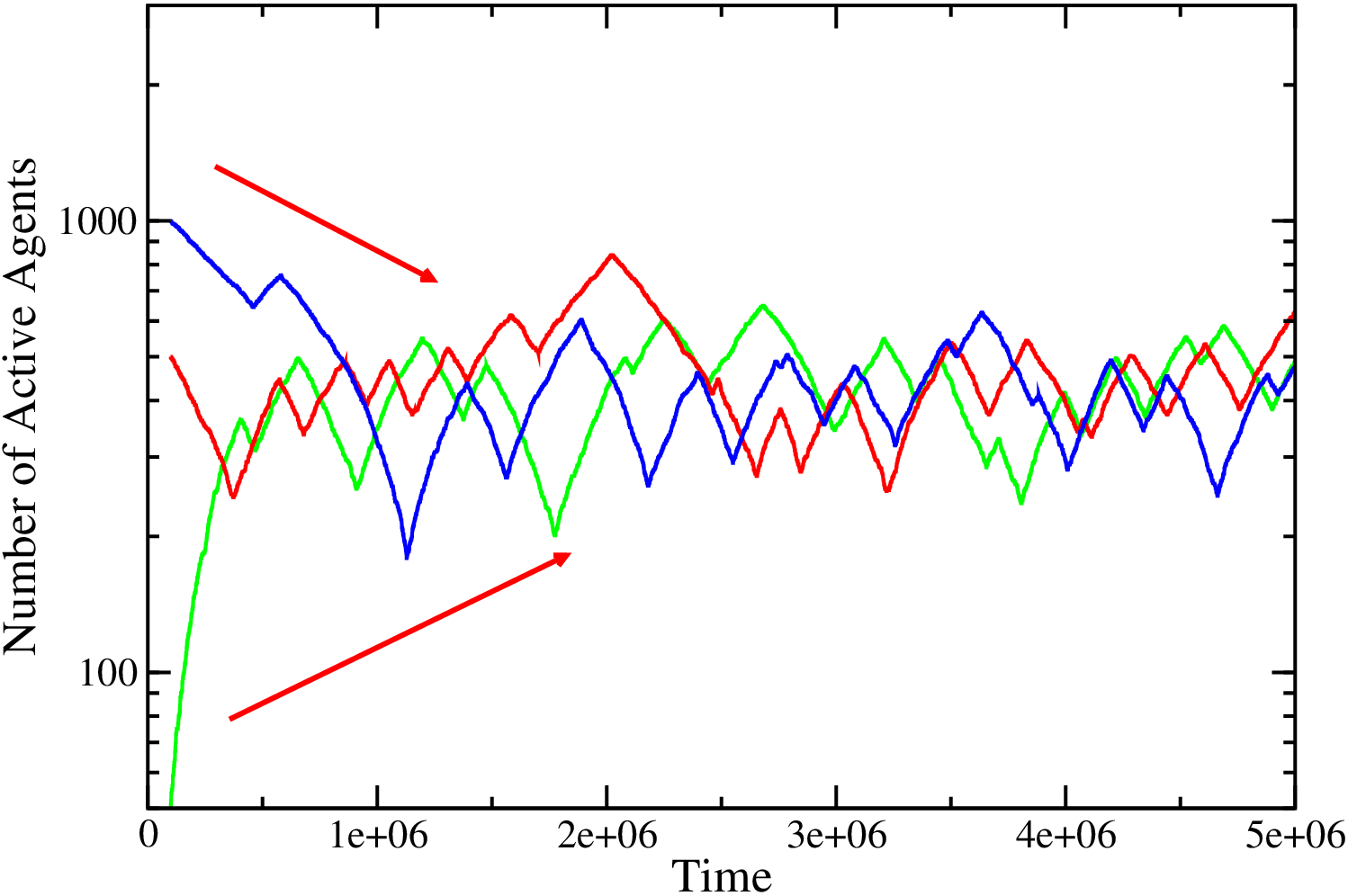}
\caption{Self-organization with only one threshold. This plot shows that the amplitude
of the region defined by $\Theta_{in}$ and $\Theta_{out}$ is not a crucial point
in the dynamics. Here we have considered a simulation where agents only compare the market volatility $\sigma$ to a unique threshold $\Theta_{SOI}$ which is chosen inside the region $\Theta_{out}$-$\Theta_{in}$.
We have plotted the self-organizations phenomenon for three different starting numbers of agents such as in Fig.~\ref{fig:soi} 
($N=50$ green line, $N=500$ red line and $N=1000$ blue line).
We can see that the results are almost the same than the ones obtained with the two thresholds dynamics. }
\label{fig:1threshold}
\end{center}
\end{figure}
Of course the interval $\Theta_{in}-\Theta_{out}$ is arbitrary and needs to be fixed.
Actually we are going to see that this is not a crucial point in order
to obtain the self-organization of the market.
In fact we have considered a situation in which agents enter or exit the market
by looking only to one suitable threshold $\Theta_{SOI}$.
Also in this case the system self-organizes in the intermittent case 
characterized by an intermediate number of agents.
The only difference with the two-thresholds dynamics, as one can see in Fig.~\ref{fig:1threshold},  is that 
the number of agents continues to have relatively large fluctuations, which resembles periodic
oscillations, even in the self-organized state.

\section{Self-organization with risk-scared agents}
\label{sec:4}

The threshold mechanism could be apparently problematic because  it may be argued that investors 
could be scared by a too fluctuating market~\cite{pagano}. 
However, this problem can easily clarified by the analysis of fluctuations at different time scales. 
The price movement which we interpret as a positive signal for the agents' strategy corresponds to the
volatility at relatively long time scale.
On the other hand a large volatility at a shorter time scale would induce a high risk on such a strategy.
In this section we consider how this problem may affect the self-organization mechanism.
We are going to see  that
the introduction of this more complex and realistic scenario in the model does not change the essential elements
of the self-organization phenomenon.

\subsection{Small scale volatility threshold}

\begin{figure}[b!]
\begin{center}
\vspace{1cm}
\includegraphics[scale=0.40]{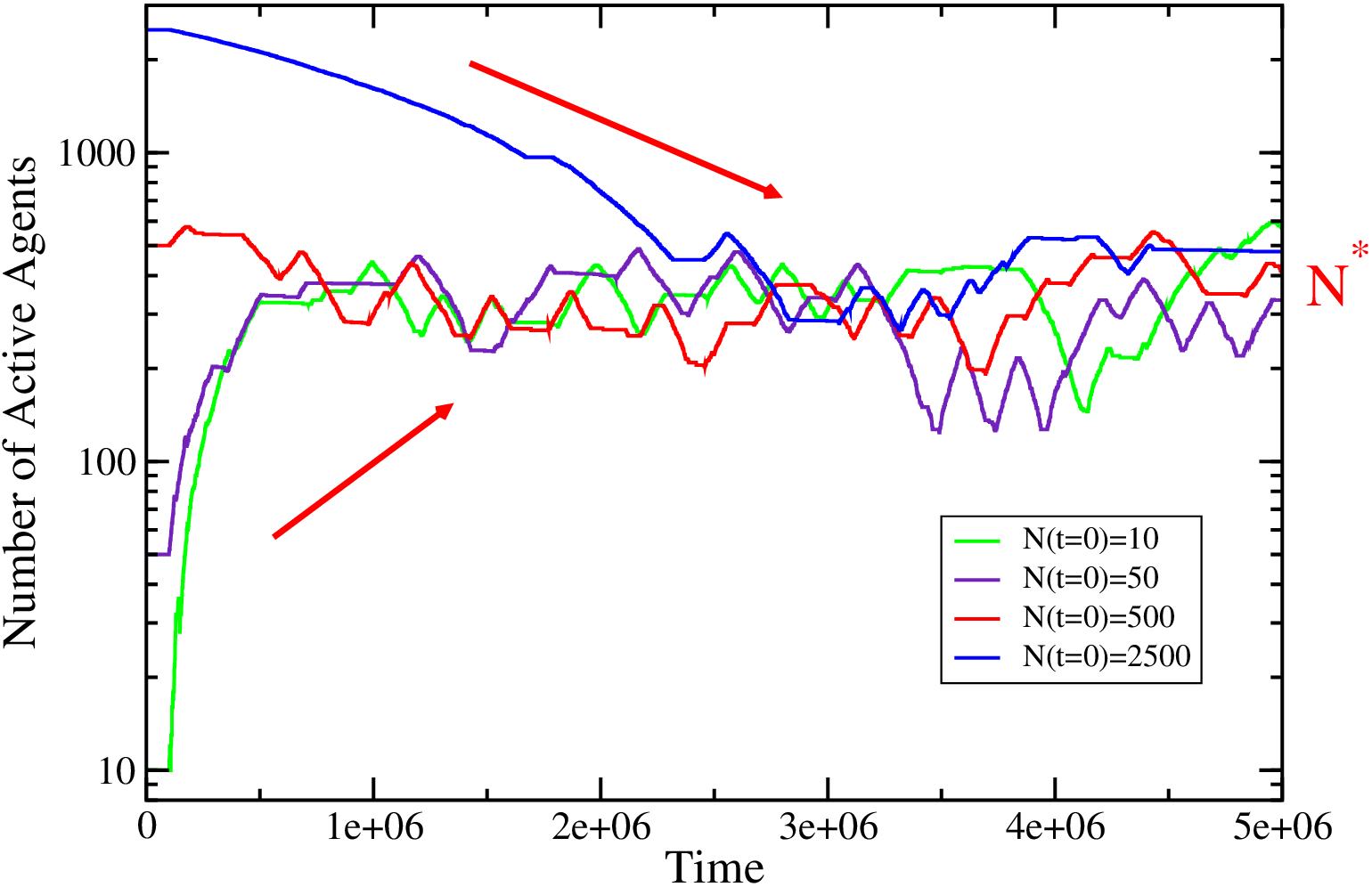}
\caption{Self-Organization with risk-scared agents.
The introduction of a small-scale threshold $\Theta_s$ does not change the results of 
Fig.~\ref{fig:soi}, as in that case, the system self-organizes into the quasi-critical state $N^*$ independently on the starting number of agents. The unique effect introduced by $\Theta_s$ is a slight asymmetry between the rise and the decrease of the number of agents $N$. }
\label{AO2}
\end{center}
\end{figure}
\begin{figure}[htbp]
\begin{center}
\includegraphics[scale=0.40]{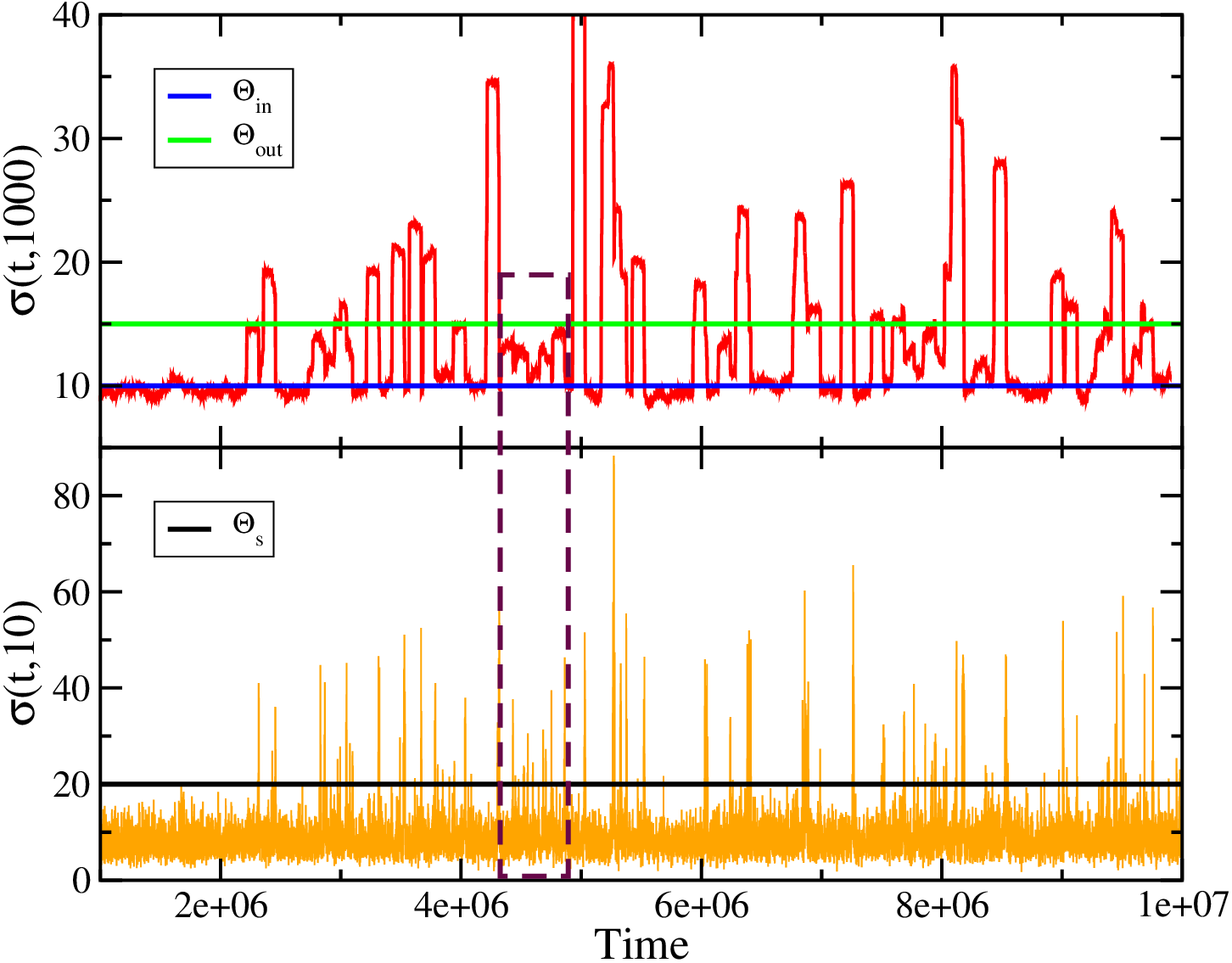}
\caption{Analysis of the volatility on different time scales. 
Large fluctuations of $\sigma(t,T_1)$ does not necessary imply large fluctuations of $\sigma(t,T_2)$ and vice-versa as it can be seen in the highlighted region. In fact while $\sigma(t,T_2)<\Theta_{in}$ we have at the same time that $\sigma(t,T_1)>\Theta_{s}$.}
\label{vol10_1000}
\end{center}
\end{figure}
\begin{figure}[t!]
\begin{center}
\includegraphics[scale=0.40]{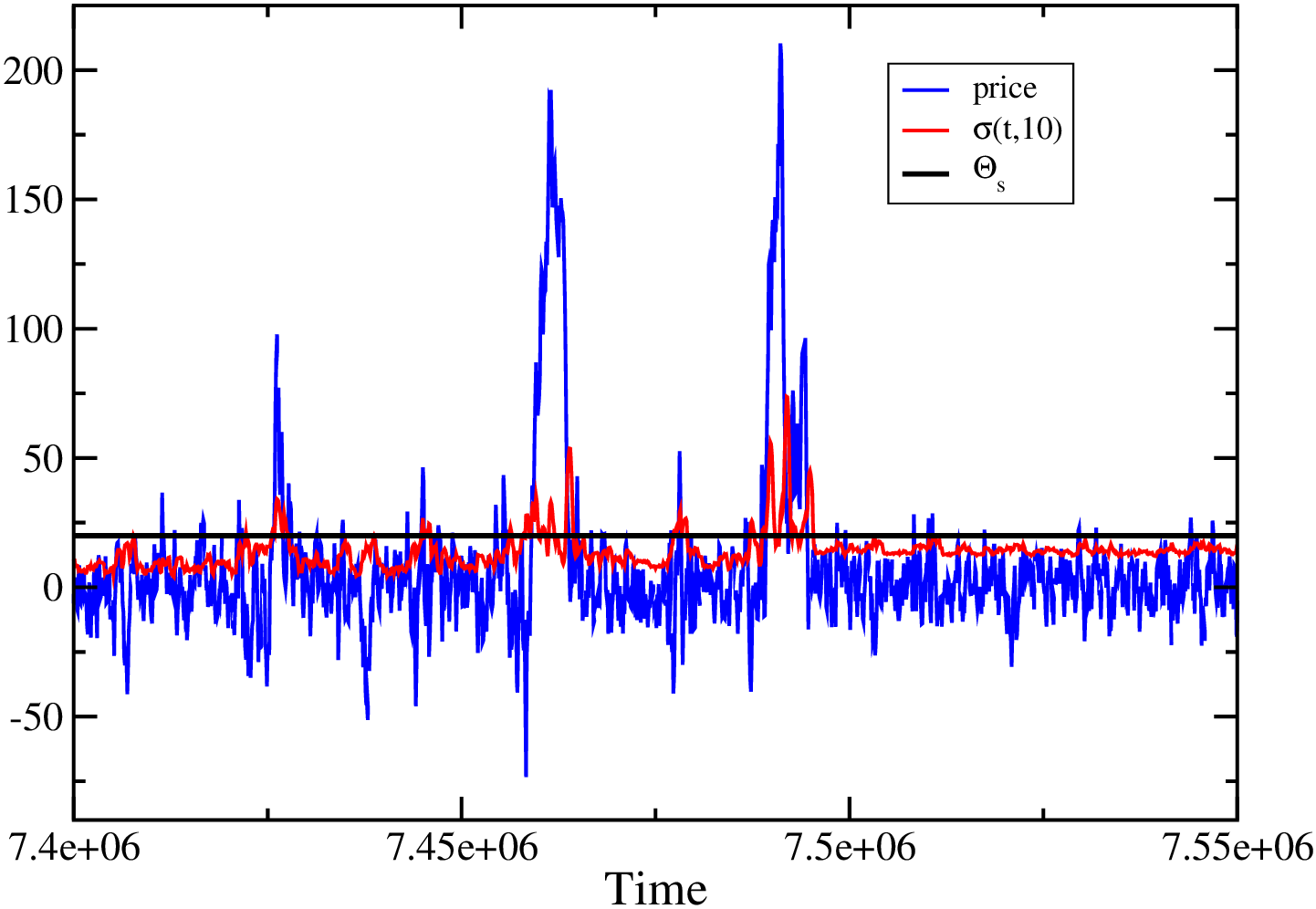}
\caption{Analysis of the short-scale volatility threshold.
We see that  the condition $\sigma(t,T_1)>\Theta_s$ is fulfilled when the price makes very large fluctuations and grows (or drops) very quickly on the scale of $T_1$.}
\label{azione}
\end{center}
\end{figure}

We use the same estimator $\sigma(t,T)$ introduced in section \ref{sec:3.1} (Eq.~\ref{eq:sigma}).
Since we want that agents look at fluctuations on two time horizons, at each time step $t$ they now have to evaluate fluctuations $\sigma(t,T)$ for two different values of $T$ that we call $T_1$ and $T_2$ corresponding respectively to the small time scale and to the large time scale. We set $T_2=1000$ as in the previous section and we choose $T_1=T_2/100$. 
\\
The fear of a too volatile market at a short time scale can be represented by the new threshold $\Theta_s$.
If $\sigma(t,T_1)>\Theta_s$ the agent will consider the situation as dangerous and she will exit the market 
with a certain probability.
If the agent is inactive and the 
previous
condition is fulfilled she will not enter in the market. Instead if the opposite condition is true (i.e $\sigma(t,T_1)<\Theta_s$) the agents compare the long time scale fluctuations $\sigma(t,T_2)$ with the thresholds $\Theta_{in},\,\Theta_{out}$ and enter/exit according to the same scheme of Eq.~\ref{eq:2thresh}. In Fig. \ref{AO2} we show the same analysis of Fig.~\ref{fig:soi} and we can see that, independently on the starting number of agents, the system tends to the quasi-critical state (i.e. $N^*$) with the SF as in the previous section for a suitable choice of the thresholds $\Theta_{in},\,\Theta_{out},\,\Theta_{s}$.
The unique effect introduced by $\Theta_s$ is a slight asymmetry between the rise and the decrease of the number of agents $N$.
The value of  $\Theta_{s}$ we adopt (Fig.~\ref{vol10_1000}) is quantitative of the same order of  $\Theta_{in}$ and 
 $\Theta_{out}$, only it corresponds to shorter time scales.
 \\
It is also interesting to note that the presence of large fluctuations on the scale of $T_1$ does not imply large fluctuations on the scale of $T_2$ or vice-versa as Fig. \ref{vol10_1000} points out. In fact we can see in the highlighted region that, while $\sigma(t,T_2)$ is smaller than $\Theta_{in}$, $\sigma(t,T_1)$ is instead usually larger than $\Theta_{s}$.
To conclude this section we report in Fig. \ref{azione} the price behavior and the fluctuations $\sigma(t,T_1)$, the small scale mechanism is active only when the price has large fluctuations
on the small scale.
\section{Conclusions}
\label{sec:5}

In this paper we have presented a critical discussion of  the self-organization phenomenon in a market model
dynamics.
We have considered a minimal ABM able to reproduce market SF
as playground to analyze the self-organization phenomenon.
In this model the dynamics strongly depends on the number of agents one considers.
For a finite intermediate number of agents one obtains an intermittent dynamics 
which leads to the SF.
In the thermodynamic limit of many agents the price fluctuations decrease and no intermittence
is observed in the dynamics. The result is a super stable market where the SF are not recovered.
Also in the limit of very few agents the SF disappear the price fluctuations being too
high.
The basic question is why real markets self-organize in the quasi critical region with the intermittent dynamics and the SF.
We propose a simple mechanism which triggers this self-organization in a population of agents
which can enter or exit the market.
The rule followed by the agents to decide to enter or leave the market is based on the idea that 
the price must undergo some appreciable movements to appear appealing for speculative 
traders. On the contrary a super stable market with the price slowly fluctuating around a fundamental
value does not display much profit opportunity and agents are not interested in.
We have studied the robustness of the self-organization with respect to variations of the threshold parameters.
The result is that it is not possible to force the system to self-organize in a state without the SF.
\\
We also consider the fact that agents may be  discouraged to enter a risky, noisy market.
This requires the analysis of fluctuations at long time scale (price movements) and short
time scale (risky volatility).
The result is that the introduction of these additional realistic effects does not appreciably modify
the self-organization phenomenon.
We believe to have characterized in a reasonably, realistic and general scheme the self-organization leading to the SF
in terms of the agents' strategies to enter or exit the market.
This is a new concept, usually neglected in ABM, which, in our opinion, should instead be considered
in the attempts to understand the origin of the SF in economic time series.

\bibliographystyle{unsrt} 
\bibliography{SOI}

\end{document}